\documentstyle[manuscript,aps,epsfig]{revtex}
\begin{document}
\flushbottom

\widetext
\draft
\title{Probing an extended region of $\Delta m^2$ with rapidly
oscillating $^7$Be solar neutrinos}
\author{A. J. Baltz}
\address{
Physics Department,
Brookhaven National Laboratory,
Upton, New York 11973}
\date{\today}
\maketitle
\def\thepage{\arabic{page}}
\makeatletter
\global\@specialpagefalse
\ifnum\c@page=1
\def\@oddhead{Draft\hfill To be submitted to Phys. Rev. D}
\else
\def\@oddhead{\hfill}
\fi
\let\@evenhead\@oddhead
\def\@oddfoot{\reset@font\rm\hfill \thepage \hfill}
\let\@evenfoot\@oddfoot
\makeatother

\begin{abstract}
If $^7$Be solar neutrinos can be observed in real time experiments, then an
extended region of $\Delta m^2$ can be probed by a proper analysis of the
rapidly changing phase of vacuum oscillations due to the eccentricity of the
earth's orbit about the sun.  For the case of maximal vacuum mixing, a kind of
Fourier analysis of expected data for one year's time could uniquely pick out
$\Delta m^2$ if it lies in the region $ \sim 10^{-10} -
6 \times 10^{-9}$(eV)$^2$.
\end{abstract}

\makeatletter
\global\@specialpagefalse
\def\@oddhead{\hfill}
\let\@evenhead\@oddhead
\makeatother
\nopagebreak
\section{Introduction}
In a previous work it was argued\cite{bgg} that with maximal vacuum mixing
there is agreement, with minor modifications, between extant observations of
solar neutrinos and predictions by the standard solar model
(SSM)\cite{bu88,bp92,bp95,bp98}.  The maximal vacuum mixing case considered
was that in which
the phase of neutrino oscillations coming from the sun is averaged, leading
to $50\%$ of the neutrinos arriving at the earth as electron neutrinos.  
As a result of this averaging, while $\sin^2 2 \theta$ was assumed to be
maximal (equal to one), $\Delta m^2$ was not determined and taken to lie in the
approximate range $10^{-9}<\Delta m^2<<10^{-3}$ with an exclusion of the
approximate range $3 \times 10^{-7}<\Delta m^2<10^{-5}$ for maximal
mixing\cite{bks} due to the lack of an observed day-night effect in the
SuperKamiokande data\cite{fuk}.  

On the other hand, the recent first
results of the SNO measurement of charged current interactions produced by
$^8$B neutrinos\cite{sno}, taken in combination with the elastic scattering
result of the Super-Kamiokande collaboration\cite{fuk}, indicate that only
about one third of the
neutrinos arriving at the earth from the sun are electron neutrinos, with the
other two thirds being $\mu$ or $\tau$ neutrinos.  Oscillation into sterile
neutrinos now seems relatively unlikely from the SNO result.

While at first glance this comparison seems to make maximal vacuum mixing
less likely,
a global analysis of the SNO result with the other solar neutrino experiments,
chlorine\cite{hom}, Super-Kamiokande\cite{fuk}, and gallium\cite{gal,sag} has
led to the conclusion that ``the CC measurement by SNO has not changed
qualitatively the globally allowed solution space for solar neutrinos, 
although the CC measurement has provided dramatic and convincing evidence
for neutrino oscillations and has strenghened the ths case for active
oscillations with large mixing angles\cite{bah01}.''  Furthermore, global
analyses\cite{bah01}\cite{fog01}
do not completely exclude solutions to the solar neutrino problem in
the mass region $ 10^{-10} < \Delta m^2 <  10^{-8}$ for maximal (or near
maximal) mixing.  In the following, the time varying phase of oscillating
$^7$Be neutrinos is investigated as a possible method to discover
(or exclude) a solution of the solar neutrino problem in that mass region.

In the mass region $4 \times 10^{-11} < \Delta m^2 <  10^{-9}$ there are
so-called ``just-so'' vacuum solutions of the solar neutrino problem,
where the phase of the oscillation
of $^8$B neutrinos coming from the sun is not completely
averaged\cite{hata,gr}.
Recall also\cite{kp2,gra}, that there is a large change in the $^7$Be
electron neutrino flux over the year in the $^8$B ``just-so'' region
due to the change in phase of order $\pi / 2$ in a year
brought about by the $\pm 1.67\%$ yearly orbital
variation from the mean distance of the sun to the earth.
As will be shown in the following, when phase averaging due to the
temperature of the sun and phase damping due to the MSW effect are
considered, it turns out that phase variation in $^7$Be neutrinos should
be observable for $\Delta m^2$ in the range from about $10^{-10}$
up to about $6 \times 10^{-9}$ (eV)$^2$.  This observable
range of $\Delta m^2$ via $^7$Be neutrinos turns out to be in approximate
agreement with a previous analysis by de~Gouv\^ea, Friedland, and
Murayama\cite{gouv}.

\section{Oscillations: thermal averaging; MSW damping}
There is a low energy region of solar neutrinos dominated by the
nearly monenergetic 862 KeV line from electron capture
by $^7$Be in the sun.  It is the purpose of the Borexino\cite{borex}
experiment, soon to come on line, to measure these neutrinos in real time.
Feasability studies are also being carried out for
other experiments to measure $^7$Be neutrinos in real time such as
LENS\cite{lens} and HELLAZ\cite{hellaz}.

If the $^7$Be neutrinos were truly monoenergetic then the number of neutrinos
detected via electron scattering in an experiment like Borexino (normalized to
unity for no oscillations) would take the following form for vacuum
oscillations
\begin{equation}
R(\phi,\Delta m^2)  = 1 - 0.79 \sin^2 2 \theta
\sin^2 { \pi \Delta m^2  L (\phi) \over (0.862) 0.00248},
\end{equation}
where $\theta$ is the vacuum mixing angle, $\Delta m^2$ is expressed in
(eV)$^2$, the $\mu$ or $\tau$ neutrino scattering relative to electron neutrino
scattering at 0.862 MeV is 0.21\cite{bah}, and
\begin{equation}
L(\phi) = {1 - \epsilon^2 \over 1 + \epsilon \cos{\phi}} 1.496 \times 10^8,
\end{equation}
the distance from the Earth to the center of the sun (in km.), which
varies through the year due to the eccentricity $\epsilon = .0167$ of the
Earth's orbit
about the sun.  Note that since we take the number of neutrinos detected
as a function of $\phi$, the phase of the earth in its orbit about the sun,
rather than of the time of the year there is no
$1 / L^2$ seasonal variation in $R$; it is canceled by the Jacobian in going
from time as an independent variable to $\phi$ as an independent variable
(Kepler's second law).

It has been pointed out by Pakvasa and Pantaleone\cite{pp} that the $^7$Be
energy line is thermally broadened not only by the spread in nuclear
velocities (Doppler broadening) but also by the solar temperature of the 
approximately 80\% of the capture electrons that come from the continuum.
We makes use of the published table of Bahcall \cite{bah1} which was obtained
by convoluting both these sources of thermal spreading to obtain the energy
profile of the 862 keV $^7$Be solar neutrino shown in Figure 1.  Note that
the distribution $w_t(x)$ is asymmetric in shape and plotted as a function of
$x = (E - 0.862)/0.862$ in MeV.
 
Expressing $\Delta m^2$ in $10^{-8} (eV)^2$ one obtains 
\begin{equation}
R(\phi,\Delta m^2) = 1 - 0.79 \sin^2 2 \theta \int dx\ w_t(x)  
\sin^2 \biggl ( { 2198 \Delta m^2  \over (1 + .0167  \cos{\phi})
(1 + x)}\biggr ) ,
\end{equation}
or 
\begin{equation}
R(\phi,\Delta m^2) = 1 - 0.395 \sin^2 2 \theta 
\biggl [1 - \int dx\ w_t(x)  
\cos \biggl ( { 4396 \Delta m^2  \over (1 + .0167  \cos{\phi}) (1 + x)
} \biggr ) \biggr ].
\end{equation}
For clarity and convenience we will retain these constants explicitly and
express $\Delta m^2$ in units of $10^{-8} (eV)^2$ unless otherwise
specified for the rest of this paper.

Since $x$ is  constrained to contribute to Eq.(9) only when it
is much smaller than unity, one may set $1/(1 + x)$ equal to $1 - x$
and carry out the integration to obtain
\begin{eqnarray}
R(\phi,\Delta m^2)& = &1 - 0.395 \sin^2 2 \theta 
\biggl [1 - I_c \biggl ( { \Delta m^2  \over 1 + .0167  \cos{\phi} }
\biggr )
\cos \biggl ( {4396 \Delta m^2  \over 1 + .0167  \cos{\phi} }
\biggr ) \\ \nonumber
&& -  I_s \biggl ( { \Delta m^2  \over 1 + .0167  \cos{\phi} } \biggr )
\sin \biggl ( {4396 \Delta m^2  \over 1 + .0167  \cos{\phi} }
\biggr ) 
\biggr ],
\end{eqnarray}
with
\begin{equation}
I_c ( u ) =\int dx\ w_t(x) \cos  ( u x ) ,
\ \ \ \ \ \ \ \ I_s ( u ) =\int dx\ w_t(x) \sin  ( u x ). 
\end{equation}

In additon to temperature damping of the the oscillations there
is a second factor that we may call ``MSW damping''.  With maximal mixing, the
phase averaged rate of electron neutrinos does not depend on whether an MSW
transition has taken place.  However, if an MSW transition has taken place in
the sun, then the maximally mixed neutrino emerges in the form of a pure mass
eigenstate (i.e. $A_1 = 0 , A_2 = 1$ in Eq.(A5) of Appendix A).  Although the
pure mass eigenstate $| \nu_2 \! >$ is half electron neutrino and half other flavor
neutrino,
there is no interference from Eq.(A5) and thus no oscillation.  The probability
of remaining an electron neutrino remains one half without variation
in the vacuum from the sun to the earth.  In contrast, pure vacuum oscillations
with maximal mixing (no MSW transition) leads to equal parts of each mass
eigenstate; the neutrinos oscillate from pure electron neutrino to pure other
flavor neutrino on the path from sun to earth.  However the phase averaged
probability of an electon neutrino reaching the earth is still one half.
Guth, Randall, and Serna\cite{grs} have pointed out the relevance
of this difference for matter oscillations in the earth: there can be
a day-night effect, even for the case of maximal mixing if there has been
an MSW transition in the sun.  Appendix A comprises a short digression
on this point.
The treatment in Appendix A assumes phase averaging over the distances
involved, due to the larger $\Delta m^2$ values that would come into play
in a possible day night
effect.  Here we are interested in the phase of the vacuum
oscillation, since that is our signal.    

The onset of MSW conversion in the sun with larger $\Delta m^2$ can be
investigated numerically by utilizing a piece of computer code adapted from
a previous investigation\cite{abjw}.  The rate of $^7$Be electron neutrinos
emerging from the sun (again normalized to unity for no oscillations)
takes the form
\begin{equation}
R(\Delta m^2)  =  A\ +\ B \cos{2 \pi \Delta m^2  X  \over (0.862) 0.00248},
\end{equation}
with $X$ the distance from the surface of the sun plus some constant.
For maximal mixing: $A = 0.5$, and $B$ is 0.5 for vacuum oscillations but
$B$ vanishes for complete adiabatic conversion. 

The top panel in Figure 2 shows how the magnitude of the
oscillation for maximum mixing is reduced with increasing
$\Delta m^2$ even though the phase averaged mixing remains a constant.
The filled circles are the values of 2$B$ calculated numerically for
$\sin^2 2 \theta = 1$.  The solid line through these circles approximates 2$B$
by $\exp[-1.583 (\Delta m^2)^2]$.  Except for the small oscillations
beyond $\Delta m^2 = 10^{-8}$, this ``MSW damping'' is well represented by
the gaussian factor.
The solid line at .5 represents $A$ for maximal mixing.  The dotted lines are
the corresponding quantities 2$B$ and $A$ for $\sin^2 2 \theta = 0.9$,
just for comparison.

Incorporating ``MSW damping'' Eq.(6) then becomes
\begin{eqnarray}
R(\phi,\Delta m^2)& = &1 - 0.395 \sin^2 2 \theta  \\ \nonumber
&&\times \biggl [1 - \exp(-1.583(\Delta m^2)^2) 
\ \biggl \{ 
I_c \biggl ( { \Delta m^2  \over 1 + .0167  \cos{\phi} }
\biggr )
\cos \biggl ( {4396 \Delta m^2  \over 1 + .0167  \cos{\phi} }
\biggr ) \\ \nonumber
&& \ \ \ \ \ \ \ \ \ \ \ \ \ \ \ \ \ \ \ \ \ \ \ \ \ \ \ \ \ \ \ \ \ \ \ \ 
+ I_s \biggl ( { \Delta m^2  \over 1 + .0167  \cos{\phi} } \biggr )
\sin \biggl ( {4396 \Delta m^2  \over 1 + .0167  \cos{\phi} }
\biggr ) \biggr \}
\ \biggr ].
\end{eqnarray}
This is the expression that we use in the calculations to follow.

Eq. (8) may also be written in the form
\begin{eqnarray}
R(\phi,\Delta m^2)& = &1 - 0.395 \sin^2 2 \theta  \\ \nonumber
&&\times \biggl [1 - \exp(-1.583(\Delta m^2)^2) 
\  
I_b \biggl ( { \Delta m^2  \over 1 + .0167  \cos{\phi} }
\biggr )
\cos \biggl ( {4396 \Delta m^2  \over 1 + .0167  \cos{\phi} } - \delta
\biggr ) 
\ \biggr ],
\end{eqnarray}
where
\begin{equation}
\delta = \arctan { I_s \over I_c},
\end{equation}
and
\begin{equation}
I_b = I_c \sec {\delta}. 
\end{equation}
For the purpose of illustration we ignore the $\cos{\phi}$ dependence
in the temperature damping factor and consider $I_b ( \Delta m^2 )$.
The bottom panel of Figure 2 shows $I_b ( \Delta m^2 )$ as the
short-dashed line
and repeats the gaussian MSW damping factor from the above panel as the
long-dashed line.  The solid line is the product of the two, the overall
damping factor including temperature spreading and MSW damping.
It is clear that there is a complete damping out of the oscillations
at $\Delta m^2 \sim 10^{-8}$
(eV)$^2$, and that this broadening averages out the phase of the oscillations
at higher values of $\Delta m^2$.
  
Finally one should note that the source broadening is insignificant because
of the following.
It turns out that the SSM density\cite{bp95} of $^7$Be neutrinos
produced as a function of the solar radius is very close to a gaussian
function of the sun's radius,
\begin{equation}
w_s(r)=\sqrt{\beta \over \pi} e^{-\beta r^2},
\end{equation}
where $\beta =  13.166 \times 10^6$, and $r$ is the distance from the center
of the sun in units of the distance from the earth to the sun.  However,
as has been pointed out\cite{pant,gouv}, the oscillations effectively
start not at
the source but at the level crossing point.  For the present  maximal mixing
case the level crossing point is at surface of the sun.  The neutrinos
originating off the
sun's axis in the direction of the earth will have a slightly larger distance
to travel due to the curvature of the sun's surface.
The gaussian density Eq.(12) leads, in a very good approximation, to a source
spreading density of the form
\begin{equation}
w_s(z)=\mu e^{- \mu z},
\end{equation}
where $\mu =  1.23 \times 10^5$ or twice $\beta$ times the ratio of the sun's
radius to the mean earth-sun distance, and $z$ is the distance from the
point on the sun's surface closest to the earth toward its center.
If this small source broadening were the only
cause of damping, then by an analytical treatment paralleling that leading
to Eq.(9), one would find a source damping factor
\begin{equation}
I( \Delta m^2 ) = {1 \over \sqrt{ 1 + (4396 \Delta m^2 / \mu )^2}}
= {1 \over \sqrt{ 1 + .001277( \Delta m^2)^2 }}.
\end{equation}
The long and short dashed line in the bottom panel of Fig.(2) represents
$I( \Delta m^2 )$.  Obviously $I( \Delta m^2 )$ only starts to deviate from
unity at the rightmost part of the plot (at $\Delta m^2 = 5$). Thus, source
broadening is insignificant for our region of interest.
  
The region that we will investigate spans the range from $\Delta m^2 \sim
10^{-10}$, the ``just so'' region for $^8$B, up to $\Delta m^2 \sim 10^{-8}$,
where the broadening averages the phase.  As noted above and in Appendix A,
one might in principle begin to see a day-night effect\cite{daynight} with the
onset of MSW damping.  In fact there would be  a sizable day-night effect for
maximal mixing at $\Delta m^2 \sim 10^{-7}$\cite{rag,abjw} (the so called
``Low'' MSW solution).   

Figure 3 shows $R(\phi,\Delta m^2)$
for maximal mixing $\sin^2 2 \theta = 1$ beginning at the low end with
$\Delta m^2$ = 0.01 (again in units of $10^{-8} (eV)^2$).  Note from Eq.(8)
that the overall phase of
the cosine factor depends on $\Delta m^2$ and that this phase changes
by $180^\circ$ when $\Delta m^2$ changes by $\pi / 4396 \cong 0.000714$.
This phase sensitivity is illustrated  Figure 3 where for each panel
in addition to the curve for the labeled value of $\Delta m^2$ there are
also curves for that value plus the appropriate increments to
shift the overall phase by $90^\circ$, $180^\circ$, $270^\circ$, and
$360^\circ$.  Figure~4 shows the increasing frequency of the
oscillations of $R(\phi,\Delta m^2)$ as a function of $\phi$ 
for the $\Delta m^2$ region of $10^{-9}$ to $10^{-8}$.  Note also the
decreasing amplitude of the oscillations as they come close to being damped
out by the temperature plus source broadening and MSW damping at $0.8 \times
10^{-8}$.

\section{Oscillations: analysis of the signal}
With such a rapid oscillation period throughout the year seen especially
in Figure 4 (on the order of several months to several days), one anticipates
that for such values of $\Delta m^2$ there would
be insufficient statistics at an experiment like Borexino
for a pattern to be obvious.  However, in what follows we will investigate how
a Fourier type analysis of data from such experiments
could give evidence of a phased oscillation and thereby determine the value
of $\Delta m^2$ if it lies in this range.  Fourier analyses of $^7$Be solar
neutrino data have been previously proposed\cite{deru,fog}, but what
follows here is a somewhat different approach.

Since Borexino is a detector rather than radiochemical
experiment, it records the information on when each count was recorded and
thereby the distance of the detector to the sun $L$ incorporated as $\phi$
in Eq.(8).  We suggest analyzing data from such experiments by effectively
integrating data with a factor
$\exp{ [ i\ 4396 \Delta m_v^2 / ( 1 + .0167  \cos{\phi}) ]}$
and varying $\Delta m_v^2$ over the range  $\sim 10^{-10} - 10^{-8}$ to look
for a signal.

To test whether the $\Delta m^2$ can be determined by such a method, Monte
Carlo data sets have been simulated in the following way.  Random numbers
are generated uniformly for $\phi$ from 0 to $2 \pi$ in order to cover the
year and make use of Eq.(8).  In order to weight events according to what
would be expected from Eq.(8) with a specific $\Delta m^2$,
a second random number between 0 and 1 is then generated for each $\phi$
and a count is generated if the random number is
less than $R(\phi,\Delta m^2)$.  This is the data set: the collection of
specific angles, $\{\phi_i\}$, at which single events are recorded durin
a year.

Figure 5 shows a sample analysis of a data set generated from 15000 Monte Carlo
attempts for $\Delta m^2 = 0.3$ in our units.  The top panel shows the
expected oscillation pattern, $R(\phi,0.3)$.  From Eq.(8) one would expect
about 9075 data points to lie below
the curve from 15000 random attempts, and in fact a set of 8993 data points
$\{\phi_i\}$ were generated in this sample.  The number of data points in this
sample corresponds roughly to a year's running time at Borexino.  
For analysis one might first consider a Fourier type
transformation on the set $\{\phi_i\}$
\begin{eqnarray}
I(\Delta m^2 , \Delta m_v^2 )&=& {1 \over n} \sum_{i=1,n} \phi_i
 \exp{ \biggl (
{i\ 4396 \Delta m_v^2  \over 1 + .0167 \cos{\phi_i} } \biggr )}\\ \nonumber
&\cong & {1 \over 2 \pi} \int_0^{2 \pi} 
R(\phi,\Delta m^2) \exp{ \biggl ( {i\ 4396 
\Delta m_v^2  \over 1 + .0167  \cos{\phi} } \biggr )}  \ d \phi.
\end{eqnarray}
The solid curve in the middle panel displays $\vert I(0.3 , \Delta m_v^2 )
\vert$ and it shows a
discontinuity in pattern near $\Delta m_v^2 = 0.3$.  If there were no phase
oscillation then one would expect $I$ to approach a function that we will
call $J(\Delta m_v^2)$
\begin{equation}
J(\Delta m_v^2) = {1 \over 2 \pi}  \int_0^{2 \pi} 
\exp{ \biggl ( {i\ 4396 
\Delta m_v^2  \over 1 + .0167  \cos{\phi} } \biggr )}  \ d \phi.
\end{equation}
The dotted line in the middle panel displays $\vert J(\Delta m_v^2) \vert$.
This suggests that we subtract off the Bessel function like behavior
$J(\Delta m_v^2)$ contained in $I(\Delta m^2 , \Delta m_v^2 )$ (which
tends to obscure the signal), create a new function
\begin{equation}
K(\Delta m^2, \Delta m_v^2) = I(\Delta m^2 , \Delta m_v^2) - J(\Delta m_v^2),
\end{equation}
and use $\vert K(\Delta m^2, \Delta m_v^2) \vert$ to analyze our data set
$\{\phi_i\}$.  $\vert K(0.3, \Delta m_v^2)\vert$ is displayed in the bottom
panel.  The signal of $\Delta m^2 = 0.3$ is unambiguous.

Figure 6 shows that an unabiguous signal would be obtained with about one
year's Borexino statistics for $\Delta m^2$ in the range 0.1--0.5.  For
$\Delta m^2 = 0.6$ there is a signal at $\Delta m_v^2 = 0.6$
that might be a little ambiguous with
only one year's statistics, but it retains its shape with increasing 
statistics; a secondary peak seen at about $\Delta m_v^2 = 0.75$
with one year's statistics goes away with the higher statistics.  There is
no signal apparent
for the $\Delta m^2 = 0.8$ case, as one would expect from looking at the
corresponding curve in Figure 4.

Figure 7 shows the extraction of the signal for $\Delta m^2$ an order of
magnitude lower.  Curves correspond to values in Figure 3.  Below
$\Delta m^2 = 0.04$ this particular analysis starts to become ambiguous.
However the slow rate of variation with $\phi$ makes direct comparison with
the patterns seen in Figure 3 practical.  At $\Delta m^2 = 0.01$
the analysis is complicated by the large change in magnitude of the rate
throughout the year with a small increment in the value of $\Delta m^2$.   

\section{Discussion}
Based on the SNO result it now seem likely that the solar neutrino puzzle has
been solved.  It is not a deficiency in the standard solar model that is
being observed but new physics.  Electron neutrinos are oscillating into some
combination of $\mu$ and $\tau$ neutrinos.  Exactly how this happens is perhaps
not yet clear, whether by one of the MSW solutions or some vacuum mixing
solution.  In the previous sections of this paper it has been shown that if
the solution to the solar neutrino puzzle happens to be maximal mixing in the
mass range $\Delta m^2 = \sim 10^{-10} - 6 \times 10^{-9}$(eV)$^2$,
then a proper
analysis of a successful $^7$Be neutrino experiment should be able to
unambiguously determine $\Delta m^2$.  Not seeing a $\Delta m^2$ signal in
this mass range would elimate a region of $\Delta m^2$ for large mixing angle.

\section{Acknowledgments}
I would like to thank Chellis Chasman for reading and commenting on the
manuscript.

This manuscript has been authored under Contract No. DE-AC02-98CH10886 with
the U. S. Department of Energy. 

\appendix
\section{The day-night effect in the limit of maximal two neutrino mixing}
Guth, Randall, and Serna\cite{grs} have pointed out that there can be
a day-night effect, even for the case of maximal mixing.  What follows is
a compact explication of this point with emphasis on the limits of no
MSW and maximal MSW effect in the sun.

The general form for two mass eigenstates in two neutrino mixing is
\begin{equation}
|\nu_1 \! >={\rm cos} \theta  |\nu_e \!  > + \ {\rm sin}
\theta |\nu_x
\! >
\end{equation} and
\begin{equation}
|\nu_2 \! >=- {\rm sin} \theta |\nu_e \!  > + \ {\rm cos} \theta |\nu_x
\! > \ ,
\end{equation}
where
$|\nu_x \! >$ is presumed to be some linear combination of $|\nu_\mu \!  >$ and
$ |\nu_\tau \!   \! >$.
Conversely
\begin{equation}
|\nu_e \! >={\rm cos} \theta  |\nu_1 \!  > - \ {\rm sin}
\theta |\nu_2
\! >
\end{equation} and
\begin{equation}
|\nu_x \! >={\rm sin} \theta |\nu_1 \!  > + \ {\rm cos} \theta |\nu_2
\! > \ .
\end{equation}
In free space mass eigenstates propagate independently.  A mixed mass state
$|\nu(t) \! >$ then has the form
\begin{equation}
|\nu(t) \! >= e^{-i m_1^2 t/2E} A_1 |\nu_1 \! > + \ e^{-i m_2^2 t/2E}
A_2 |\nu_2 \! >.
\end{equation}
The probability $P_s$ that a neutrino born in the sun is an electron neutrino
when it reaches the earth is then
\begin{equation}
P_s = {\rm cos}^2 \theta P_1 + {\rm sin}^2 \theta P_2
\end{equation}
with $P_{1,2} = \vert A_{1,2} \vert^2$ the average probability of a mass one
or mass two eigenstate arriving at the earth where the phase has been averaged
by the distances involved.  Since $P_1 = 1 - P_2$ this
may also be written equivalently
\begin{equation}
P_2 = {( {\rm cos}^2 \theta - P_s) \over {\rm cos} 2 \theta }
\end{equation}
The probability $P$ that an electron neutrino born in the sun will be an
electron neutino after passing through the sun, traveling to the earth, and
then passing through the earth is simply
\begin{equation}
P = P_2 P_{2e} + (1 - P_2) (1 - P_{2e})
\end{equation}
where $P_{2e}$ is the probability that a mass 2 neutrino entering the earth
emerges at the detector as an electron neutrino.
Making use of Eq.(A7) this becomes
\begin{equation}
P = P_s + {(1 - 2 P_s) \over {\rm cos} 2 \theta } (P_{2e} - {\rm sin}^2
\theta).
\end{equation}
This expression is the Mikeyev-Smirnov
expression\cite{msw} for the day night effect, trivially transformed\cite{abjw}
to be most transparent in various limits.

The maximum value for $P_s$ occurs for vacuum oscillations
\begin{equation}
P_s = 1 - {1 \over 2} {\rm sin}^2 2 \theta
\end{equation}
and
\begin{equation}
{(1 - 2 P_s) \over {\rm cos} 2 \theta } = -{\rm cos} 2 \theta.
\end{equation} 

The minimum value for $P_s$ occurs for complete adiabatic MSW conversion to
a pure mass eigenstate $|\nu_2\! >$.  In this case from Eq.(A6) 
\begin{equation}
P_s = {\rm sin}^2 \theta
\end{equation}
and
\begin{equation}
{(1 - 2 P_s) \over {\rm cos} 2 \theta } = 1.
\end{equation}

Thus as $\sin 2 \theta$ goes to 1 (maximal mixing) there is no
day-night effect for vacuum oscillations and a maximum effect possible in the
case of complete adiabatic conversion.
\vfill\eject

\begin{figure}
\epsfig{file=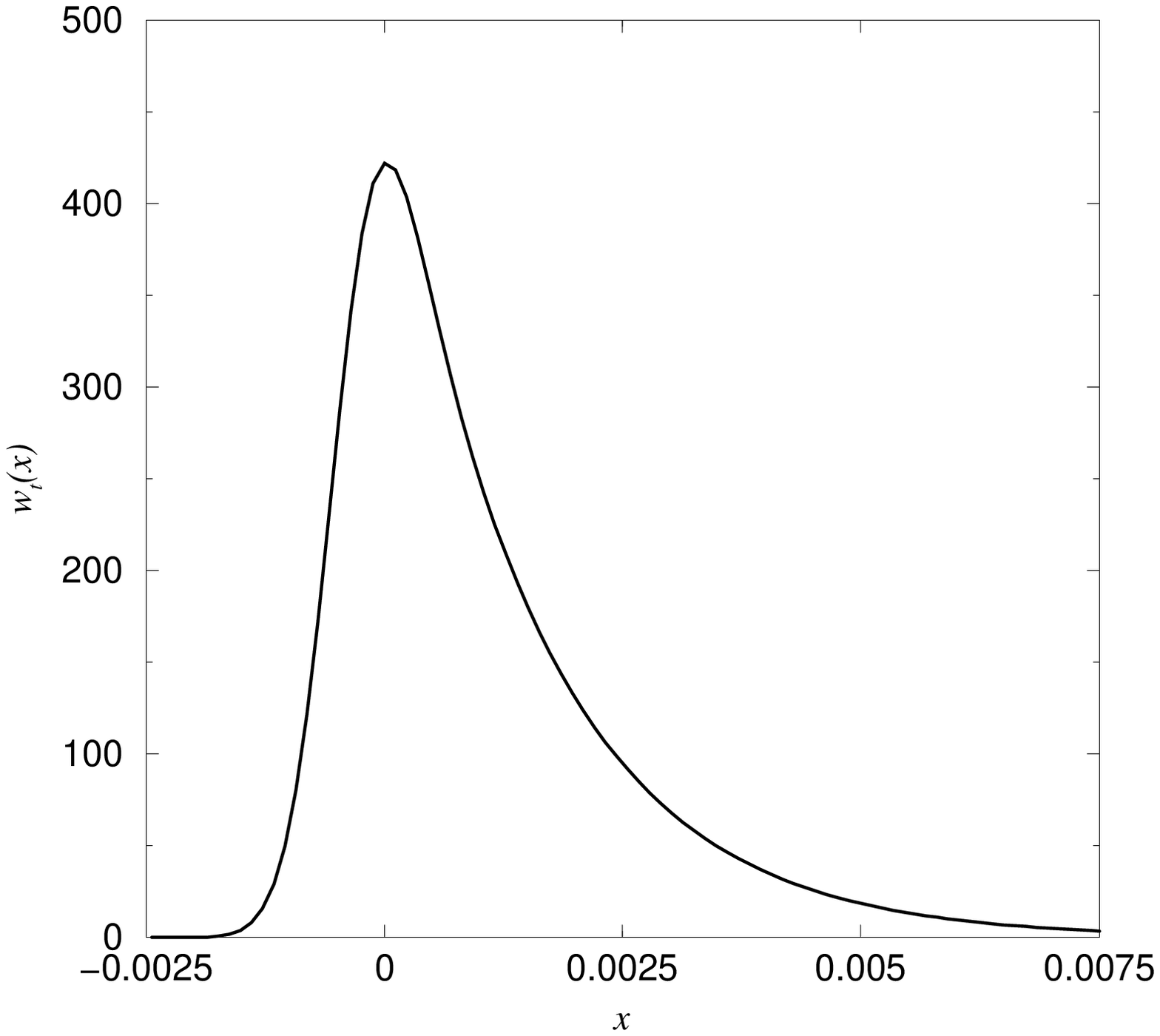,width=15cm,height=15cm}
\caption[Figure 1]{Temperature broadening of $^7$Be neutrinos originating in
the sun (see text).}
\end{figure}
\begin{figure}
\epsfig{file=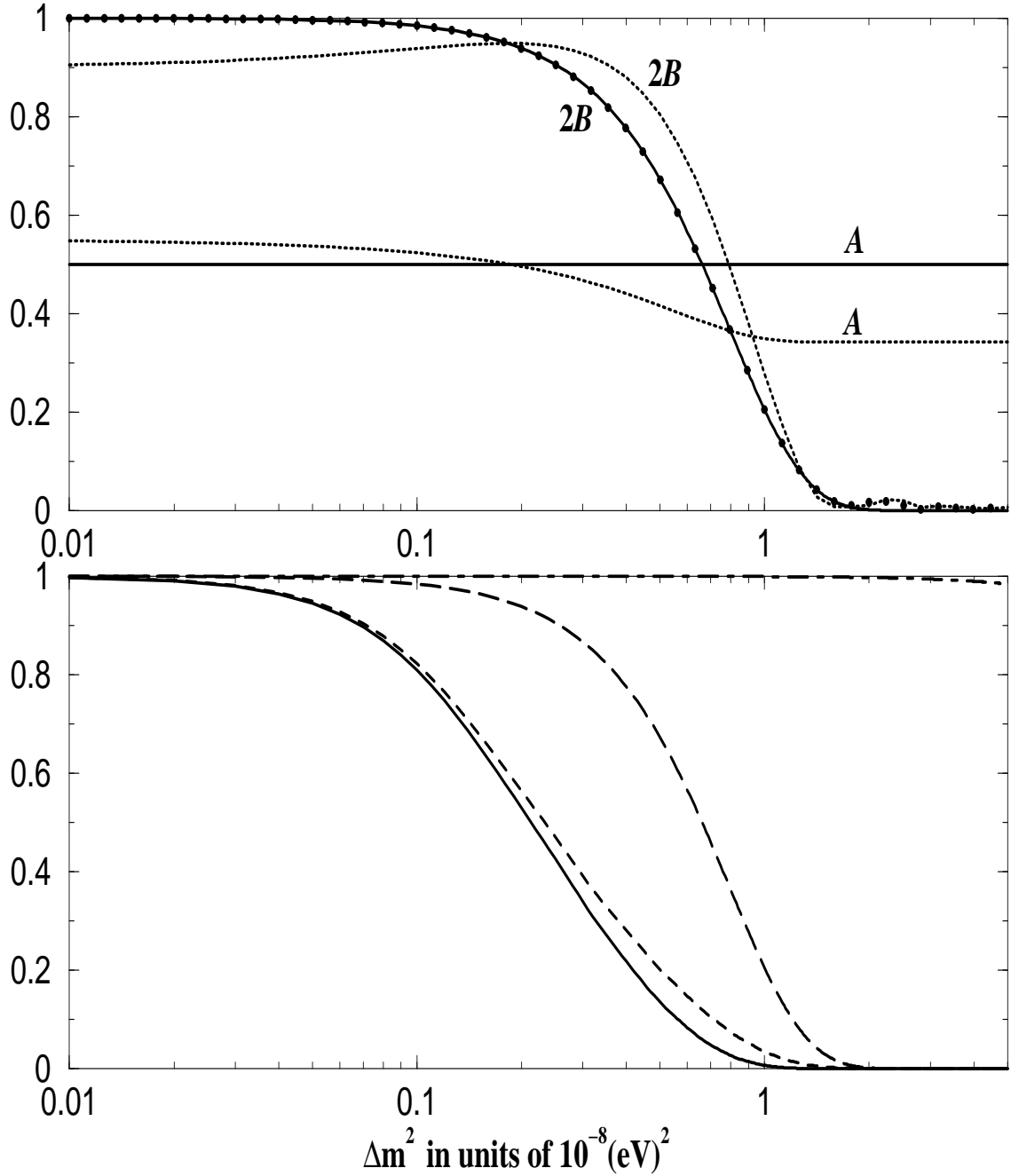,width=15cm,height=18cm}
\caption[Figure 2]{Top panel: Constant $A$ and oscillating $B$ part of electron
neutrino rate emerging from the sun for $\sin^2 2 \theta = 1$ (solid line) and
for $\sin^2 2 \theta = 0.9$, (dotted line);
Bottom panel: Total oscillation damping factor (solid line) and partial damping
factors (see text).}
\end{figure}
\begin{figure}
\epsfig{file=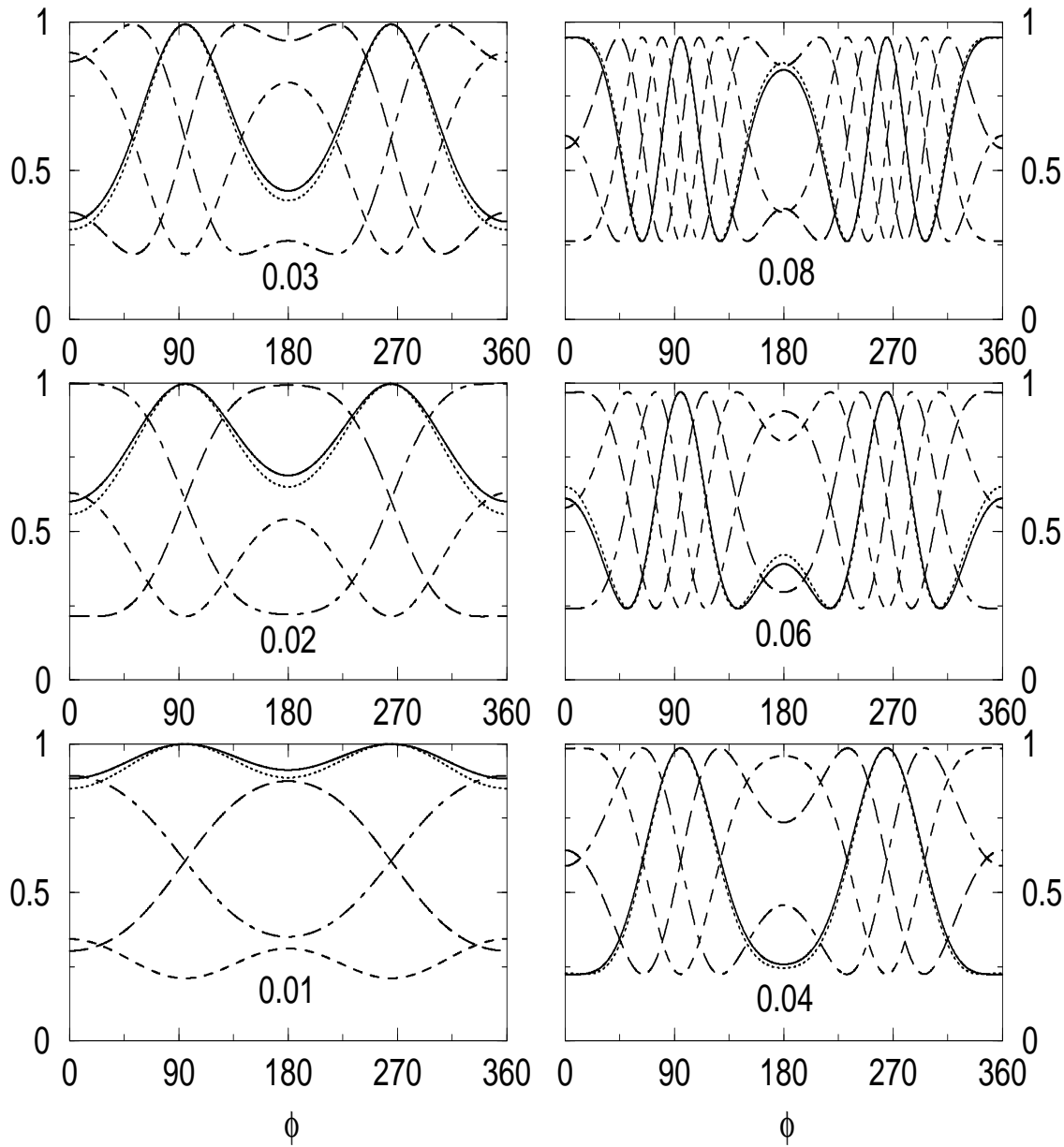,width=15cm,height=17cm}
\caption[Figure 3]{Rate of $^7$Be neutrinos detected by electron scattering
for maximal
mixing.  The number on each panel gives $\Delta m^2$ for the solid line.
The long and short dashed line is $\Delta m^2$ plus approximately 0.00035,
the short dashed line $\Delta m^2$ plus 0.0007, the long dashed line
$\Delta m^2$ plus 0.00105 and the dotted line $\Delta m^2$ plus 0.0014.}
\end{figure}
\begin{figure}
\epsfig{file=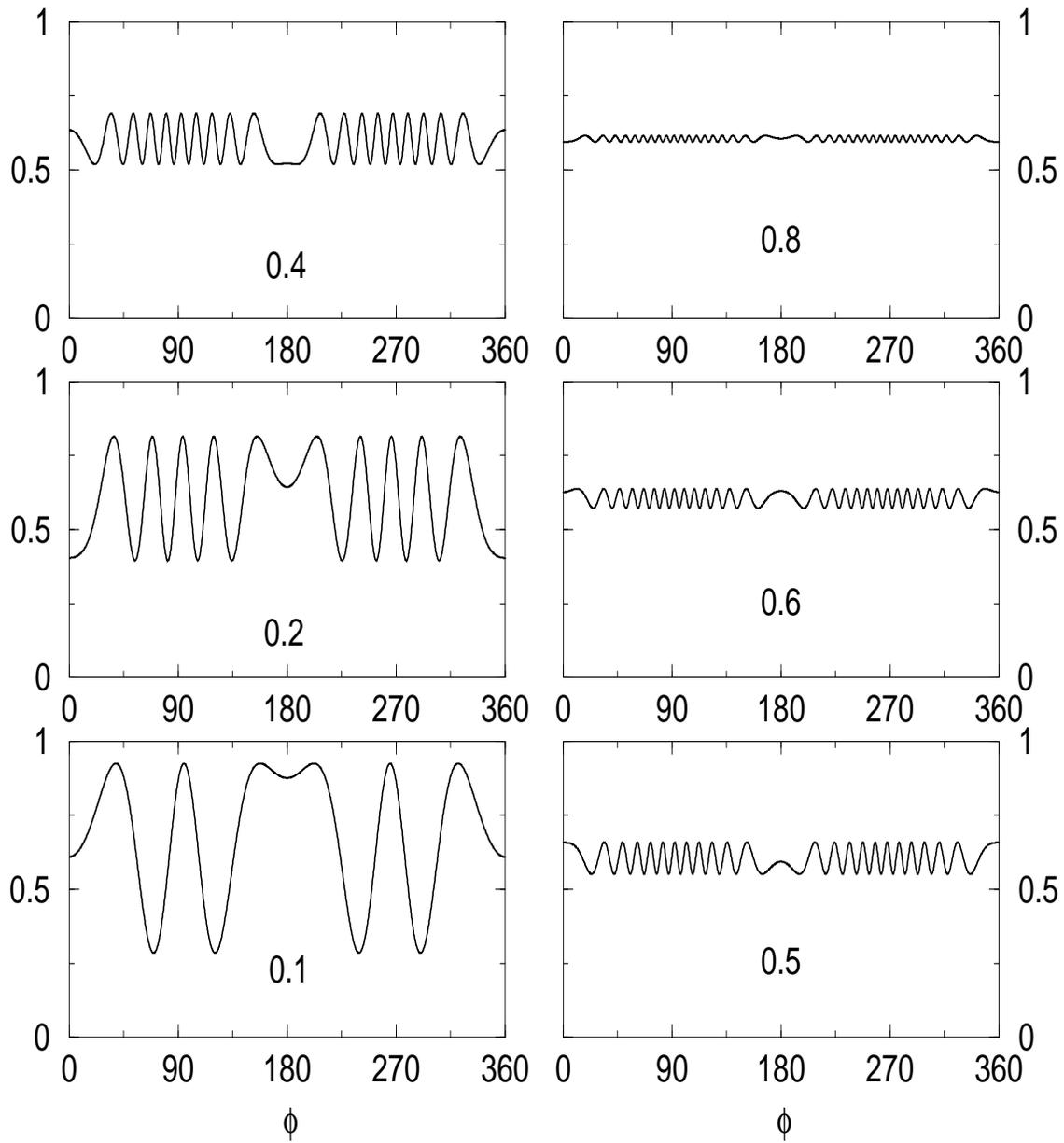,width=15cm,height=17cm}
\caption[Figure 4]{Rate of $^7$Be neutrinos detected by electron scattering
for maximal
mixing.  The number on each panel gives $\Delta m^2$ for the solid line.}
\end{figure}
\begin{figure}
\epsfig{file=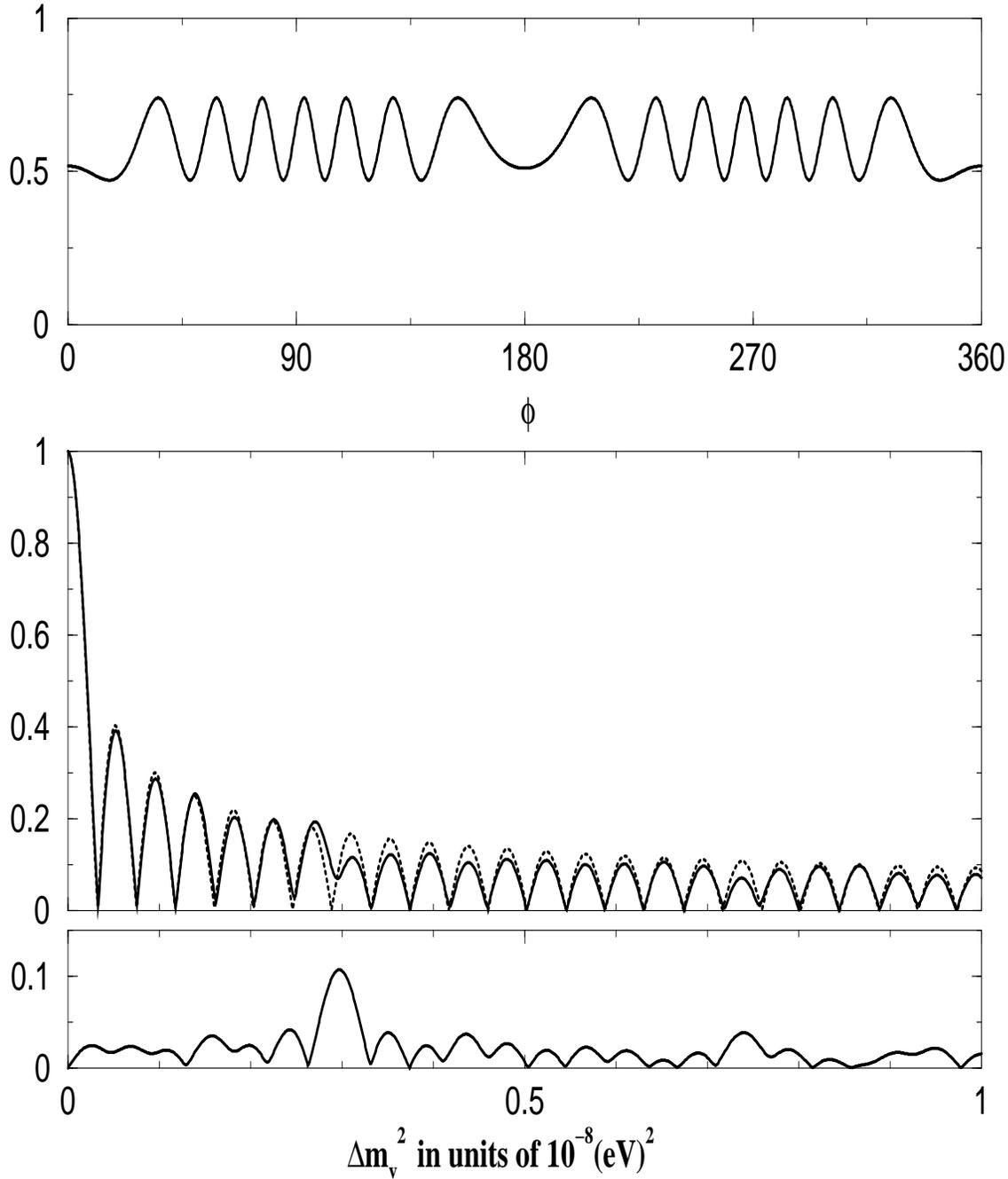,width=15cm,height=18cm}
\caption[Figure 5]{The top panel gives the rate of $^7$Be neutrinos coming
from the sun for maximal mixing and $\Delta m^2$ = 0.3.  The solid line
in the middle panel gives the absolute value of the Fourier analysis of the
distribution $\vert I(0.5 , \Delta m_v^2 )
\vert$ and the dashed line the absolute
value of the analysis of the constant average of the distribution 
$\vert J(\Delta m_v^2) \vert$.  The bottom panel shows the absolute
value of the difference of the Fourier analyses of the distribution and
the constant average of the distribution $\vert K(0.5, \Delta m_v^2)\vert$.
This is the quantity that extracts the $\Delta m^2$ signal.}
\end{figure}
\begin{figure}
\epsfig{file=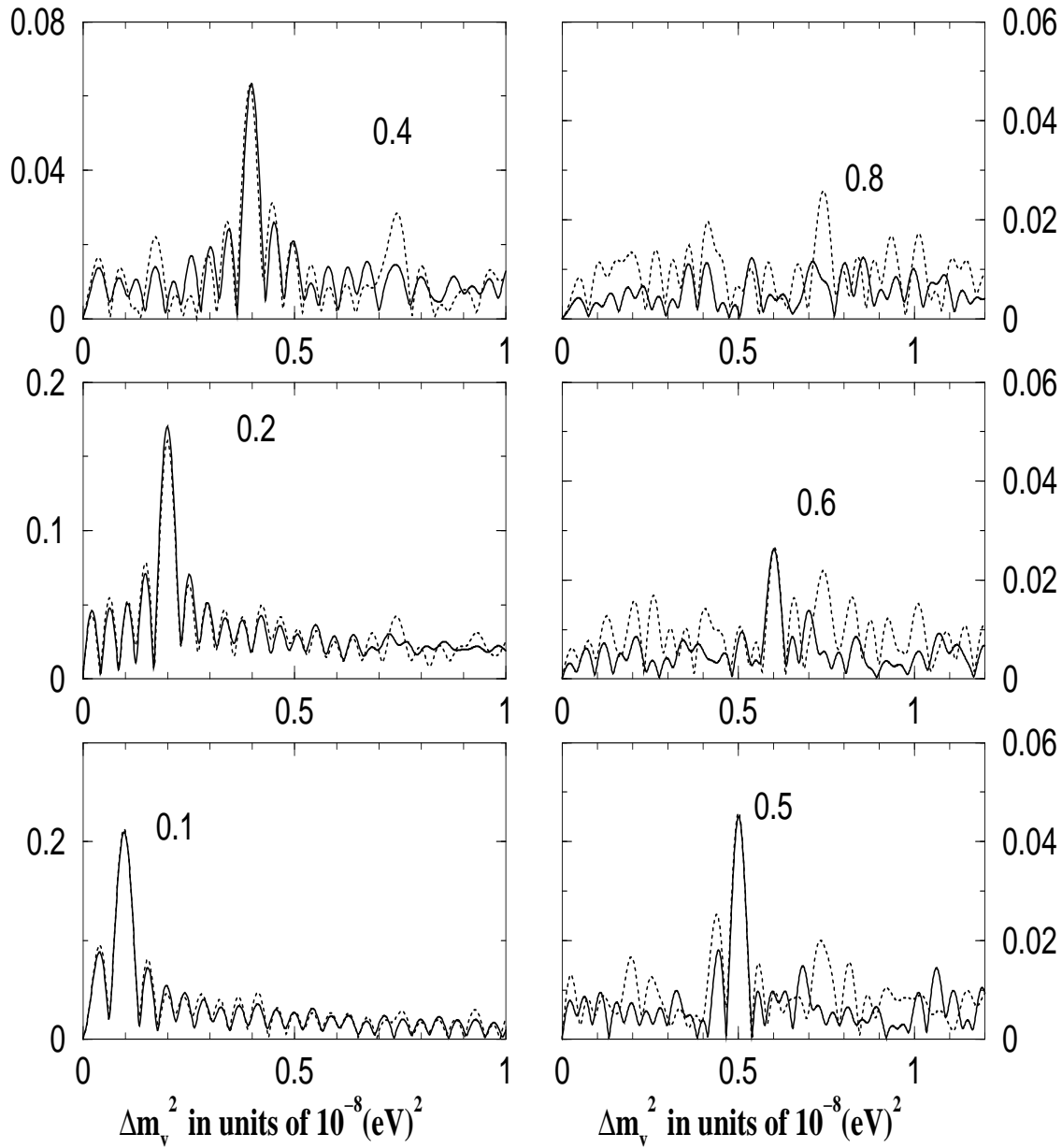,width=15cm,height=17cm}
\caption[Figure 6]{Extraction of the the signal
$\vert K(\Delta m^2, \Delta m_v^2)\vert$.  The number on each panel gives
the value of $\Delta m^2$.  The dotted line lines represents 15000 tries or
about 9000 events.  The solid lines represent four times the statistics:
60000 tries or about 36000 events.}
\end{figure}
\begin{figure}
\epsfig{file=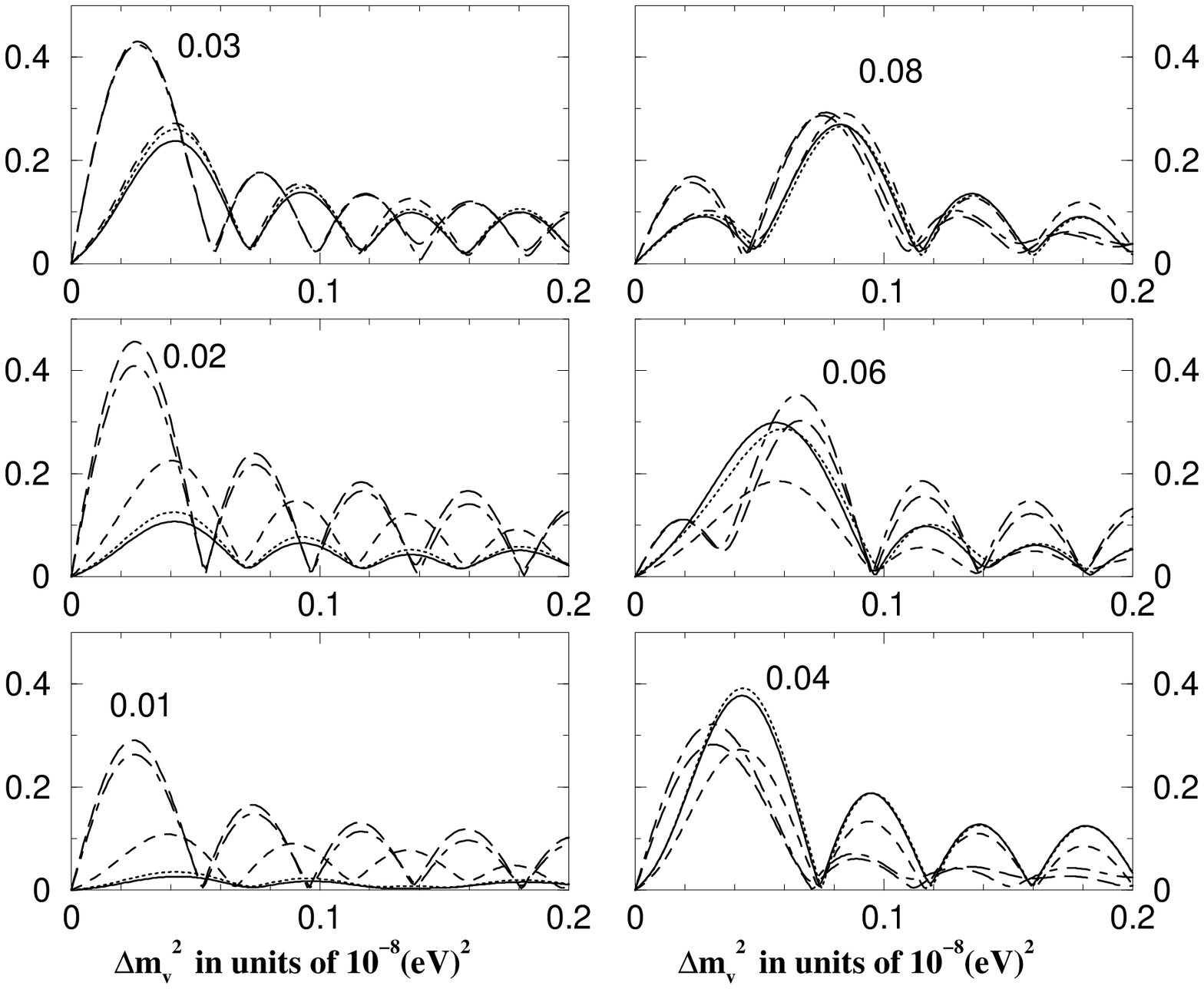,width=15cm,height=17cm}
\caption[Figure 7]{$\vert K(\Delta m^2, \Delta m_v^2)\vert$ as in Figure 6,
 but for lower values of 
$\Delta m^2$.  The location of the peak becomes less well defined with
decreasing $\Delta m^2$.  As in Figure 3 the long and short dashed line
is $\Delta m^2$ plus approximately 0.00035,
the short dashed line $\Delta m^2$ plus 0.0007, the long dashed line
$\Delta m^2$ plus 0.00105 and the dotted line $\Delta m^2$ plus 0.0014.}

\end{figure}

\begin{references}
\bibitem {bgg} Anthony J. Baltz, Alfred Scharff Goldhaber, and Maurice
Goldhaber, Phys. Rev. Lett. {\bf 81}, 5730 (1998).
\bibitem {bu88} J. N. Bahcall and R. N. Ulrich,
Rev. Mod. Phys. {\bf 60}, 297 (1988). 
\bibitem {bp92} J. N. Bahcall and M. H. Pinsonneault,
Rev. Mod. Phys. {\bf 64}, 885 (1992). 
\bibitem {bp95} John N. Bahcall, M. Pinsonneault, and G. J. Wasserburg,
Rev. Mod. Phys. {\bf 67}, 781 (1995). 
\bibitem {bp98} J. N. Bahcall, S. Basu, and M. H. Pinsonneault, Phys. Lett.
{\bf B 433}, 1 (1998)
\bibitem {bks} J. N. Bahcall, P. I. Krastev, and A. Yu. Smirnov, Phys. Rev.
{\bf D 58}, 096016 (1998).
\bibitem {fuk} Y. Fukuda et al. (Super-Kamiokande Collaboration), Phys.
Rev. Lett. {\bf 81}, 1158 (1998).
\bibitem {sno} Q. R. Ahmed, et al. (SNO Collaboration), Phys. Rev. Lett.
{\bf 87}, 071301 (2001)
\bibitem {hom} B. T. Cleveland et al., Astrophys. J. {\bf 496}, 505 (1998).
\bibitem {gal} W. Hampel et al. (GALLEX Collaboration), Phys. Lett. {\bf B
447}, 127 (1999).
\bibitem {sag} J. N. Abdurashitov et al. (SAGE Collaboration), Phys. Rev.
{\bf C 60}, 055801 (1999).
\bibitem {bah01} John N. Bahcall, M. C. Gonzalez-Garcia, and 
Carlos Pe\~na-Garay,\hfill\break
arXiv:hep-ph/0106258.
\bibitem {fog01} G. L. Fogli, E. Lisi, D. Mantanino, and A. Palazzo,
arXiv:hep-ph/0106247.
\bibitem {hata} N. Hata, and P. G.  Langacker, Phys. Rev. {\bf D 56}, 6107
(1997).
\bibitem {gr} James M. Gelb and S. P. Rosen, Phys. Rev. {\bf D 60}, 011301
(1999).
\bibitem {kp2} P. I. Krastev and S. T. Petcov, Nucl. Phys. {\bf B 449}, 605
(1995).
\bibitem {gra} James M. Gelb and S. P. Rosen, arXiv:hep-ph/9908325. 
\bibitem {gouv} Andr\'e de Gouv\^ea, Alexander Friedland, and Hitoshi
Murayama, Phys. Rev. {\bf D 60}, 093011 (1999).
\bibitem {borex} Borexino Collaboration, G. Alimonti et al.,
arXiv:hep-ex/0012030. 
\bibitem {lens} M. Cribier for the LENS collaboration, Nucl. Phys.
{\bf B} (Proc. Suppl.){\bf 87}, 195, (2000).
\bibitem {hellaz} P. Gorodetzky, A. de Bellefon, J. Dolbeau, T. Patzak,
P. Salin, A. Sarrat, J. C. Vanel, Nucl. Phys.
{\bf B} (Proc. Suppl.){\bf 87}, 506, (2000).
\bibitem {bah} J. N. Bahcall, {\it Neutrino Astophysics}, Cambridge University
Press (1989).
\bibitem {pp} Sandip Pakvasa and James Pantaleone, Phys. Rev. Lett. {\bf 65},
2479 (1990).
\bibitem {bah1} John N. Bahcall, Phys. Rev. {\bf D 49}, 3923 (1994).
\bibitem {grs} Alan H. Guth, Lisa Randall, and Mario Serna, JHEP {\bf 9908},
018 (1999).
\bibitem {abjw} A. J. Baltz and J. Weneser, Phys. Rev. {\bf D 50} 5971 (1994).
\bibitem {pant} J. Pantaleone, Phys. Lett. {\bf B 251}, 618 (1990).
\bibitem{daynight}A. J. Baltz and J. Weneser, Phys. Rev. {\bf D 35} 528 (1987);
{\bf D 37} 3364 (1988); E. D Carlson, Phys. Rev. {\bf D 34}, 1454 (1986);
J. Bouchez, M. Cribier, W. Hampel, J. Rich, M. Spiro, and D. Vignaud,
Z. Phys. C {\bf 32}, 499 (1986).
\bibitem {rag} R. S. Raghavan, A. B. Balantekin, F. Loreti, A. J. Baltz,
S. Pakvasa, and J. Pantaleone, Phys. Rev. {\bf D 44}, 3786 (1991).
\bibitem {deru} A. De Rujula and S. Glashow, CERN-TH 6655/92 (1992).
\bibitem {fog} G. L. Fogli, E. Lisi and D. Mantanino, Phys. Rev. {\bf D 56}
4374 (1997).
\bibitem {msw} S. P. Mikheyev and A. Yu. Smirnov, in {\it New and Exotic
Phenomena,} Proceedings of the Moriond Workshop, Les Arc, Savoie, France, 1987,
edited by O. Fackler and J. Tran Thanh Van (Editions Fronti\`ieres,
Gif-sur-Yvette, France, 1987), p. 405.
\end{references}
\end{document}